\documentclass[12pt]{article}

\usepackage{a4wide}
\usepackage{amsmath,amssymb}
\usepackage{graphicx}
\usepackage{dcolumn}
\usepackage{bm}
\usepackage{hyperref}

\newcommand{\pd}{{\partial}}
\newcommand{\rd}{{\mathrm{d}}}
\newcommand{\ee}{{\mathrm{e}}}
\newcommand{\ri}{{\mathrm{i}}}

\newcommand{\la}{\langle}
\newcommand{\ra}{\rangle}

\begin{document}

\title{Boltzmann Entropy for Quantum Field Systems}

\author{Kyo Yoshida\\
  \textit{Division of Physics, Faculty of Pure and Applied Sciences,} \\
  \textit{University of Tsukuba, 1-1-1 Tennoudai,} \\
  \textit{Tsukuba, Ibaraki 305-8571, Japan}
  }


\date{}


\maketitle

\begin{abstract}
  A way to construct Boltzmann entropy, i.e., the entropy as a function
  of a microscopic pure state, for quantum field systems is proposed.  
  Operators that shift the field in wavevector space are used in
  the construction.  By employing an assumption
  that some terms emerging due to the shift are negligible
  in the thermodynamic limit,
  it is shown that, 
  for almost all states in the ensemble of pure states corresponding to
  a thermodynamic state, the value of the proposed Boltzmann entropy
  coincides with that of the thermodynamic entropy for the
  thermodynamic state.  For general self-interacting fields,
  the Boltzmann entropy evolves with time under Hamiltonian dynamics,
  so that it is capable of characterizing the thermalization of
  isolated quantum field systems.
\end{abstract}


\section{Introduction}
The statistical mechanics for thermal equilibrium states in
classical and quantum systems are well established in the sense 
that it satisfactorily provides prescription to obtain macroscopic
thermodynamic variables, such as entropy $S$,
Helmholtz free energy $F$, and many others, from
the Hamiltonian that determines the microscopic and dynamical
nature of the system.  However, it may be said that
the general consensus on its foundations has not been reached.

Regarding the definition of thermal equilibrium in microscopic term,
there are two views which may be called the
\textit{ensemblist view} and the \textit{individualist view}, where
we borrowed the terminology from 
Ref.\cite{GoldsteinLebowitzTumulkaZanghi2010}.  In the former
view, a system is in thermal equilibrium when it is a mixed state
(or an ensemble of states) that is close to the canonical
or microcanonical mixed state (or ensemble) , whereas in
the latter view, a pure state (or a point in phase space) can be
in or out of thermal equilibrium depending on the state.

The ensemblist view has traditionally prevailed, 
presumably because of its concise formalism.
Once the canonical, microcanonical or some other ensemble
corresponding to a thermal equilibrium state is defined from
the Hamiltonian of the system, all corresponding thermodynamic
variables can be derived through suitable statistical calculations
with respect to the ensemble.  For example, in classical systems,
the entropy $S$ is given by the form of Gibbs entropy 
$S=-\int_\Gamma \rd x \rho(x) \ln \rho(x)$ where $\Gamma$ is the phase
space and $\rho(x)$ is a probability density function on $\Gamma$
representing the ensemble.  In the case of a canonical ensemble 
with inverse temperature $\beta:=(k_{\mathrm{B}}T)^{-1}$ where 
$k_{\mathrm{B}}$ is the Boltzmann constant, 
the probability density function is given by 
$\rho^{\mathrm{can},\beta}(x)=\exp(-\beta H(x))/Z(\beta) (x\in\Gamma)$
where $H(x)$ is the Hamiltonian and $Z(\beta)$ is the partition function.  
In quantum systems, the ensemble is represented by
the density operator $\rho$; for example, the canonical density operator
is given by $\rho^{\mathrm{can}}(\beta)=\exp(-\beta H)/Z(\beta)$ with
the Hamiltonian $H$ being an operator, and the entropy is given by
the form of von Neumann entropy $S=-\mathrm{tr}[\rho \ln\rho]$.  

On the other hand, the individualist view has its basis 
in the concept of typicality.  Consider an ensemble of (pure) 
states of a large classical or quantum system and a function of state. 
Here, we say that the function satisfies the typicality with 
respect to the ensemble if the value of the function 
is almost same for almost all states in the ensemble.  
The state is said to be typical when the value of the function is 
very close to the ensemble average and it is said to be atypical when 
the value of the function largely deviates from the average.  
In the individualist view, thermodynamic variables are 
functions of (pure) states and they are expected to satisfy 
the typicality with respect to ensembles corresponding to 
thermal equilibrium, say microcanonical ensembles or 
canonical ensembles.  Then, we may identify the values of 
the
thermodynamic 
variables at typical states, 
which we interpret as thermal equilibrium states, 
with the ensemble average of those.  
In the individualist view, the physical objects are 
individual (pure) states and ensembles are just working tools 
for computing thermodynamic variables efficiently.  
It is actually shown in 
Refs.~\cite{Sugita2006,Sugita2007,PopescuShortWinter2006,GoldsteinLebowitzTumulkaZanghi2006} that, in a large quantum system, 
physical quantities associated with a small subsystem satisfy 
typicality with respect to the microcanonical ensembles of 
the large system.  Thus, the individualist view has been gaining 
theoretical support.  

As far as thermodynamic variables in thermal equilibrium are concerned, 
all the outcome are the same for the both views and there may be little 
practical reason to distinguish them.  Significant difference 
between the two views emerges in the case of nonequilibrium processes, 
especially in thermalization in isolated systems.  
Note that the Gibbs entropy and the von Neumann entropy do not 
evolve with time under the Hamiltonian dynamics of isolated systems.  
Therefore, these entropies are not capable of characterizing 
thermalization.  In the individualist view, thermalization can be 
understood as follows:  When a nonequilibrium state, that is 
an atypical state, is chosen as the initial state, 
it would develop with time into thermal states because 
they are typical in the ensemble, and the system substantially 
never comes back to the original nonequilibrium 
state nor develops into other nonequilibrium states 
because they are so rare.  

Recently, thermalization in isolated macroscopic quantum systems 
has become a quite active research topic
(see, e.g., Ref.~\cite{Tasaki2016} and references cited therein). 
The individualist view plays an important role in the recent studies 
and there is a growing consensus that the nature of thermodynamic 
equilibrium states and thermalization can be analyzed through 
a single typical pure state and its time evolution 
\cite{RigolDunjkoOlshanii2008,SugiuraShimizu2012,SugiuraShimizu2013,IyodaKanekoSagawa2017}.  Rigol, Dunjko and Olshanii\cite{RigolDunjkoOlshanii2008} 
demonstrated that a pure state of an isolated system of 
hard-core bosons with additional weak nearest-neighbor repulsions 
on a two-dimensional lattice do thermalize 
under Hamiltonian dynamics by monitoring relaxation of 
the central component of the marginal momentum distribution.  

From thermodynamic point of view, entropy is the function 
that characterizes the thermalization processes.  It generally 
increases under adiabatic thermalization processes.  
So it is of interest to consider the entropy in the individualist
view, that is, entropy defined for microscopic state, or (pure) state.  
It should satisfy typicality and its ensemble 
average should coincide with the entropy defined for the ensemble, 
i.e., the Gibbs entropy or the von Neumann entropy.  
The concept of entropy as a function of microscopic state 
is originated by Boltzmann and here we call such entropy 
the Boltzmann entropy.  (See e.g., Ref.~\cite{Lebowitz2008} for a review
of the Boltzmann entropy. See also the introduction of
Ref.~\cite{GoldsteinLebowitzTumulkaZanghi2019arxiv} for
a similar use of the terminology of Gibbs and Boltzmann entropy.)
Boltzmann introduced a decomposition 
of the phase space into disjoint subregions in the construction of
the Boltzmann entropy for classical systems.
For quantum systems, the pioneering work on the Boltzmann entropy 
was given by von Neumann\cite{vonNeumann1929}. 
(Hereafter, Ref.~\cite{vonNeumann1929} 
is referred to as VN29. See Ref.~\cite{GoldsteinLebowitzTumulkaZanghi2010} 
for a review of VN29 from a modern perspective. )  
In VN29, the Boltzmann entropy is defined 
by using the abstract formalism of the decomposition of the total Hilbert
space.  Although the existence of many appropriate decompositions of
the total Hilbert space is guaranteed, the prescription of the appropriate 
decompositions for specific systems was not given.  

The aim of this paper is to construct a Boltzmann entropy, 
a function of pure state, for quantum systems on a lattice, 
or systems of quantum field, and to show that it satisfies 
some properties that are desirable for the entropy.  
We restrict the system to quantum field in order to give
the explicit construction of the Boltzmann entropy.  
The present study is inspired by the author's previous work
on the construction of entropy as a function of state in classical 
field systems\cite{Yoshida2015}.  Following the previous work, 
the construction of the Boltzmann entropy is considered in the 
wavevector space and operators that shift the field 
in the wavevector space are used. 

Recently, \v{S}afr\'{a}nek, Deutsch and 
Aguirre\cite{SafranekDeutschAguirreR2019,SafranekDeutschAguirre2019} 
proposed a decomposition of the total Hilbert space based on 
a coarse-graining in position space to construct a Boltzmann 
entropy for systems of many quantum particles 
(hereafter, Refs.~\cite{SafranekDeutschAguirreR2019,SafranekDeutschAguirre2019} 
are referred to as \v{S}DA19).
Our study shares interest and aim with \v{S}DA19 to some extent but 
the two studies developed to construct different types of Boltzmann entropy.  
Comparison between the two studies is given in Sec. \ref{sec:discussions}.  
Very recently, there is an attempt to define not c-number valued but
an operator valued Boltzmann entropy in
Ref.~\cite{GoldsteinLebowitzTumulkaZanghi2019arxiv}.
However, we do not expand the scope of this paper to include
the operator valued Boltzmann entropy.  

This paper is organized as follows:  We start with the general 
formalism of constructing the Boltzmann entropy
in Sec. \ref{sec:formalism}.  
We then introduce a normal distribution model for ensemble of 
pure states in Sec. \ref{sec:normaldistributionmodel}. Next,
we give the setting of the quantum field systems in Sec.
\ref{sec:quantumfield}. 
After these preparations, we construct a Boltzmann entropy 
for the quantum field systems and examine its properties 
by using the normal distribution model for ensemble of pure states
in Sec. \ref{sec:boltzmannentropy}. 
We conclude with some discussion of the results in
Sec. \ref{sec:discussions}.  

\section{General formalism}
\label{sec:formalism}
We first consider the properties that the Boltzmann entropy 
$S(\psi)$, i.e., entropy as a function of (microscopic, pure) state $\psi$,
should satisfy, if it is formulated, for both in classical and quantum systems. 
Let $X$ be a set of thermodynamic variables that specifies 
a thermodynamic equilibrium state.  For example, 
$X=(U,V,N)$ or $X=(\beta,V,N)$, where
$U$ is the internal energy, $V$ is the volume, $N$ is the number 
of particles and $\beta$ is the inverse temperature.  
In conventional statistical mechanics, 
a thermodynamic equilibrium state is modeled by 
a probability density function (PDF) $P(\psi)$ on 
the phase space $\Gamma$ for classical systems 
and a density operator $\rho$ for quantum systems.  
In the next section, we introduce a model 
PDF $P(\psi)$ on the Hilbert space $\mathcal{H}$
which corresponds to a given density operator $\rho$, so that 
a thermodynamic equilibrium state $X$ both in classical and
quantum systems can be described by a PDF on
the state space, which will be denoted by
$P^{(X)}(\psi)$.  Let the average with respect to $P^{(X)}(\psi)$ be 
denoted by
\begin{equation}
\overline{F(\psi)}^{(X)}:=\int_{\mathcal H}\rd\psi P^{(X)}(\psi) F(\psi), 
\end{equation}
for an arbitrary function $F(\psi)$ of the state $\psi$.  
The overline without superscript $(X)$ will mean the average 
with respect to an arbitrary PDF $P(\psi)$.  

So that the Boltzmann entropy $S(\psi)$ is consistent with 
the thermodynamic entropy $S(X)$ defined for thermodynamic state $X$, 
the following conditions should be required. 
\begin{description}
\item{(S1)} The average of $S(\psi)$ with respect to $P^{(X)}(\psi)$ is equal
to the thermodynamic entropy $S(X)$ in the thermodynamic limit,
say $V\to\infty$, i.e., 
\begin{equation}
\lim_{V\to \infty}\frac{\overline{S(\psi)}^{(X)}}{S(X)}=1. 
\end{equation}

\item{(S2)} The deviation from the mean 
$\Delta S(\psi):=S(\psi)-\overline{S(\psi)}^{(X)}$
is small in the thermodynamic limit in the sense that 
\begin{equation}
  \lim_{V\to\infty}
  \frac{\overline{\left(\Delta S(\psi)\right)^2}^{(X)}}
  {\left(S(X)\right)^2}
       =0.  
\end{equation}
\end{description}
Note that the thermodynamic limits in the above should be taken 
by fixing the intensive variables such as $N/V$, $E/V$ and $\beta$.  
By virtue of the Chebyshev's inequality, the conditions
(S1) and (S2) imply that
$S(\psi)$ is typically almost equal to $S(X)$ when $\psi$ is randomly
chosen according to the PDF $P^{(X)}(\psi)$.  

In addition to being consistent with the thermodynamic entropy, 
it is desirable for the Boltzmann 
entropy $S(\psi)$ to have some properties characterizing nonequilibrium 
states and processes.  Let $\psi(t)$ denote the time evolution of
the state with respect to the Hamiltonian $H$ and
the initial condition $\psi(t=0)=\psi$.  
If $S(\psi)$ is substantially smaller than $S(X)$ and $S(\psi(t))$
approaches $S(X)$ as $t\to \infty$, then we may interpret that
$\psi$ is an atypical nonequilibrium state and that $\psi(t)$ is
a nonequilibrium processes of thermalization.
So that such $\psi(t)$ does exist, $S(\psi)$ should
at least satisfy the following condition.
\begin{description}
\item{(S3)}
$\rd S(\psi(t))/\rd t|_{t=0}\ne 0$ for some $\psi$ with finite 
probability for finite $V$.  
\end{description}

Now we proceed to introduce the general formalism of 
the Boltzmann entropy $S(\psi)$ for both classical and quantum systems.  
We start with the formalism for classical systems.  
Let $\Gamma$ be the phase space of a classical systems.  
For every state $\psi \in \Gamma$, we assign 
a function $f(\phi;\psi)$ of the state $\phi \in \Gamma$, 
which satisfies the properties of a probability density function 
with respect to $\phi$, i.e., 
\begin{equation}
\int_{\Gamma}\rd \phi f(\phi;\psi) =1,\qquad f(\phi;\psi) \ge 0. 
\label{eq:probability}
\end{equation}
We call the function $f(\phi;\psi)$ the density function associated
with the microscopic state $\psi$.  This function $f(\phi;\psi)$ should
be distinguished from the PDF $P^{(X)}(\phi) (\phi\in\Gamma)$
associated to macroscopic thermodynamic state $X$.
Hereinafter, we use the symbol ``$f$'' for 
the density function associated with an individual microscopic state in
order to avoid confusion.  A Boltzmann entropy $S(\psi)$, i.e., 
entropy as a function of microscopic state, can be introduced as 
\begin{equation}
S(\psi)=-\int_{\Gamma}\rd\phi f(\phi;\psi) \ln f(\phi;\psi). 
\label{eq:Spsi_cl}
\end{equation}
The Boltzmann entropy $S(\psi)$ is completely determined by
the choice of the function $f(\phi;\psi)$.  
The function $f(\phi;\psi)$ should be determined so as 
the corresponding Boltzmann entropy $S(\psi)$ to satisfy 
the desirable conditions (S1)--(S3).  
A conceptual guideline to determine 
$f(\phi;\psi)$ may be given by introducing 
a notion of ``macroscopic similarity''.
For a fixed state $\psi$, assign a value of $f(\phi;\psi)$ 
for every state $\phi$ according to the degree of macroscopic 
similarity between $\phi$ and $\psi$.  That is, let
$f(\phi;\psi) >f(\phi';\psi)$ if $\phi$ is macroscopically 
more similar to $\psi$ in comparison with $\phi'$.
Conversely, if $f(\phi;\psi)$ is already defined, 
we may estimate the degree of the macroscopic similarity of 
$\phi$ to $\psi$ by the value of $f(\phi;\psi)$.  

When a decomposition of the phase space 
$\Gamma=\bigcup_{\nu} \Gamma_\nu$ 
into mutually disjoint subregions $\Gamma_\nu$ is given
and states in a same subregion may be regarded to be macroscopically 
similar to each other, 
one can define the function $f(\phi;\psi)$ as 
\begin{equation}
f(\phi;\psi)=|\Gamma_{\nu(\psi)}|^{-1} \delta_{\nu(\phi) \nu(\psi)}, 
\end{equation}
where $|\Gamma_{\nu}|$ is the volume of $\Gamma_{\nu}$, 
$\nu(\psi)$ and $\nu(\phi)$ are such that $\psi \in \Gamma_{\nu(\psi)}$, 
and $\phi \in \Gamma_{\nu(\phi)}$, respectively, and $\delta_{\nu\nu'}$ is
the Kronecker delta.
Equation (\ref{eq:Spsi_cl}) reduces to 
\begin{equation}
S(\psi)=\ln |\Gamma_{\nu(\psi)}|. 
\label{eq:Spsi_cl_decom}
\end{equation}
A decomposition of the phase space is obtained by specifying
a sequence of the energy values $U^{(\nu)} (\nu=0,1,2,\cdots )$
satisfying $U^{(0)} < U^{(1)} < U^{(2)}<\cdots$ with
$U^{(0)}$ being the ground energy and letting 
$\Gamma_{\nu}=\{\psi|U^{(\nu-1)} <H(\psi) \le U^{(\nu)}\} (\nu=1,2,\cdots)$
where $H(\psi)$ is the Hamiltonian.  Then $S(\psi)$ in
Eq. (\ref{eq:Spsi_cl_decom}) is consistent with
the Gibbs entropy for the microcanonical ensemble
$S_{\textrm{mc}}(U)$ with $U=H(\psi)$.  
Another example of decomposition is given by introducing 
a set of real functions $\{M\}(\psi)=\{M_1(\psi),\ldots,M_n(\psi)\}$ 
which corresponds to macroscopic quantities.  
Let the sequences $M_m^{(\nu_m)} (\nu_m=0,1,2,\cdots; m=1,2,\cdots,n)$ 
satisfy $M_m^{(0)} <M_m^{(2)}<\cdots$, and the subregion is
defined by 
$\Gamma_\nu=\{\psi|M_m^{(\nu_m-1)}<M_m(\psi)\le M_m^{(\nu_m)}, m=1,\cdots,n\}$
for $\nu=\{\nu_1,\cdots,\nu_m\}$.  Then $S(\psi)$ in
Eq. (\ref{eq:Spsi_cl_decom}) is nothing but 
the Boltzmann entropy $S_{\mathrm{B}}(\{M\}(\psi))$ given in
Ref.~\cite{Lebowitz2008}.

The equivalent formalism for quantum systems is given as follows.  
Let $\mathcal H$ be the Hilbert space and we assign an operator
$f(\psi)$ to every pure state $|\psi\ra \in \mathcal H$ which satisfies
the properties of density operator, i.e., 
self-adjoint, positive semidefinite, and of trace one.  
Again, we will use the symbol ``$f$'' for the density operator 
associated with individual pure state to distinguish it from 
the density operator $\rho(X)$ associated to macroscopic thermodynamic
state $X$.  
A Boltzmann entropy $S(\psi)$ is given by 
\begin{equation}
S(\psi)=
-\mathrm{tr}\left[f(\psi)\ln f(\psi)\right]. 
\label{eq:entropy_quantum}
\end{equation}
The operator $f(\psi)$ can be expressed as 
\begin{equation}
f(\psi) = \int \rd \phi\ w(\phi;\psi)
U(\phi;\psi)|\psi\ra\la\psi|
U^\dagger(\phi;\psi), 
\label{eq:rho_psi_gen}
\end{equation}
where $\phi$ is a parameter
which may have many components, 
$U(\phi;\psi)$ 
is a unitary operator labeled by $\phi$ and $\psi$, 
and $w(\phi;\psi)\ge 0$ is a weighting function 
satisfying $\int \rd\phi\ w(\phi;\psi)=1$.  
%
The unitary operator $U(\phi;\psi)$ maps
the state $|\psi\ra$ to $U(\phi;\psi)|\psi\ra$ 
and the value of $w(\phi;\psi)$ gives 
the degree of macroscopic similarity of
$U(\phi;\psi)|\psi\ra$ to $|\psi\ra$.
The Boltzmann entropy is completely determined by
$U(\phi;\psi)$ and $w(\phi;\psi)$ so that the problem is
the choice of them.  

For example, 
let us consider the case when a decomposition of
the Hilbert space $\mathcal{H}$ into mutually orthogonal
$n$ subspaces $\mathcal{H}_\nu (\nu=1,\ldots,n)$, i.e.,
$\mathcal{H}=\bigoplus_{\nu=1}^n \mathcal{H}_\nu$, where $n$
can be infinite, is given according to VN29.
Let $\mathcal{U}(\mathcal{H}_\nu)$ be
the set of unitary operators acting on $\mathcal H_\nu$ and
$\mu_\nu$ be the Haar measure on $\mathcal{U}(\mathcal{H}_\nu)$
normalized as 
$\int_{U_\nu\in \mathcal{U}(\mathcal{H}_\nu)} \rd \mu_\nu (U_\nu) 1 = 1$.  
Let $U_\nu\in\mathcal U(H_\nu)$ and let the parameter $\phi$ 
be $\{U_\nu\}=\{U_1,U_2,\ldots,U_n\}$.  We put 
$U(\phi=\{U_\nu\};\psi)=\prod_{\nu=1}^n U_\nu$ and
$d\phi\ w(\phi;\psi)=\prod_{\nu=1}^n \rd\mu_\nu (U_\nu)$.  Note that
$U(\phi;\psi)$ and $w(\phi;\psi)$ are now independent of $\psi$.  
By noting that
\begin{equation}
\int_{\mathcal{U}(\mathcal{H}_\nu)}\rd\mu_\nu(U_\nu) U_\nu A U_\nu^\dagger
=\frac{\mathrm{tr}[\mathsf{P}_\nu A\mathsf{P}_\nu]}
{\mathrm{tr}[\mathsf{P}_\nu]} \mathsf{P}_\nu +
(I-\mathsf{P}_\nu) A (I-\mathsf{P}_\nu), 
\label{eq:UAUdagger}
\end{equation}
where 
$\mathsf{P}_\nu$ is the projection operator on $\mathcal{H}_\nu$ and
$I$ is the identity operator on $\mathcal{H}$, one obtains 
\begin{align}
  f(\psi)&=\sum_\nu
\frac{\mathrm{tr}[\mathsf{P}_\nu |\psi\ra\la\psi|\mathsf{P}_\nu]}
{\mathrm{tr}[\mathsf{P}_\nu]} \mathsf{P}_\nu,\\
S(\psi)&=-
\sum_\nu
\mathrm{tr}[\mathsf{P}_\nu |\psi\ra\la\psi|\mathsf{P}_\nu]
\ln
\frac{\mathrm{tr}[\mathsf{P}_\nu |\psi\ra\la\psi|\mathsf{P}_\nu]}
{\mathrm{tr}[\mathsf{P}_\nu]} .
\label{eq:S_decom}
\end{align}

\section{Normal distribution model for the ensemble of pure states}
\label{sec:normaldistributionmodel}
As mentioned in Sec.~\ref{sec:formalism}, we need an ensemble of pure 
states, or PDF $P^{(X)}(\psi)$ on the Hilbert space $\mathcal{H}$, 
which corresponds to a given thermodynamic equilibrium state $X$ 
in the case of quantum systems.  
The density operator $\rho(X)$ that corresponds to a thermodynamic
equilibrium state $X$ can be obtained through the standard method
of microcanonical ensemble, canonical ensemble or some other
variant ensembles. In this section, we introduce a method to construct 
the PDF $P^{(X)}_{\mathrm{N}}(\psi)$ from $\rho(X)$, where
the subscript $\mathrm{N}$ stands for the normal distribution.  

For a given density operator $\rho$, 
the way to decompose it into an ensemble, i.e., a weighted sum,  
of pure states $\rho=\sum_\alpha P(\alpha)|\alpha\ra\la \alpha|$, 
is not unique.  The spectral decomposition is an evident example,
but there are many other choices.  Here, we introduce a decomposition 
which yields a multivariate complex normal distribution 
$P_{\mathrm N}(\psi)$ on the Hilbert space $\mathcal{H}$, i.e., 
\begin{align}
  \rho&=\int_{\mathcal H}\rd\psi P_{\mathrm N}(\psi)|\psi\ra\la\psi|,
\label{eq:rho_decompose}  \\
  P_{\mathrm N}(\psi)&=\frac{1}{\pi^{D} \det \rho}\exp[-\la \psi|\rho^{-1}|\psi\ra], 
\label{eq:Ppsi}
\end{align}
where $\rd\psi=\prod_{\alpha=1}^{D} 
\rd\mathrm{Re}\psi_{\alpha}\mathrm{Im}\psi_{\alpha}$,
$\psi_{\alpha}=\la \alpha|\psi\ra$, $|\alpha\ra (\alpha=1,\ldots,D)$ 
is an arbitrary orthonormal basis,
$\mathrm{det}\ \rho$ is the determinant of a matrix expression of 
$\rho$, and $D$ is the dimension of the Hilbert space 
$\mathcal{H}$.
Equation (\ref{eq:Ppsi}) implies that the components $\psi_{\alpha}$ obey
the multivariate complex normal distribution with
the means $\overline{\psi_{\alpha}}=0$ and the 
covariances $\overline{\psi_{\alpha}\psi_{\beta}^*}=\la \alpha|\rho|\beta\ra$.  
Then, it is easily checked that Eq. (\ref{eq:rho_decompose}) holds by 
computing the matrix elements $\la \alpha|\bullet|\beta\ra$ of the
both sides.  
When $\rho$ has $0$ eigenvalue for an eigenvector $|\alpha\ra$,
$P(\psi)$ in Eq. (\ref{eq:Ppsi}) is not well-defined.
In such a case, we may put $\rho|\alpha\ra=\epsilon_\alpha|\alpha\ra$
with $\epsilon_\alpha >0$ in Eq. (\ref{eq:Ppsi})
and the right-hand-side of Eq. (\ref{eq:rho_decompose}) is safely
obtained by taking the limit $\epsilon_\alpha \to +0$.  
For the case of $D=\infty$, we can first restrict the Hilbert space 
to be a finite-dimensional subspace $\mathcal H' \subset \mathcal H$ 
and consider the projection of $|\psi\ra$ onto $\mathcal H'$ and 
then increase the dimension of $\mathcal H'$ to infinity.  
Equation (\ref{eq:rho_decompose}) implies that $\rho$ is
equivalent to the ensemble 
of pure states $|\psi\ra$ with the probability density function 
Eq. (\ref{eq:Ppsi}).  

Although we have $\overline{\la\psi|\psi\ra}=1$, individual 
states $|\psi\ra$ in the ensemble are not normalized, 
$\la \psi|\psi\ra \ne 1$, in general.  However, 
if 
$(\Delta \la\psi |\psi\ra)^2:=
\overline{(\la\psi|\psi\ra -\overline{\la\psi|\psi\ra})^2}$ is 
small enough, 
then $|\psi\ra$ is normalized as $\la\psi |\psi\ra=1$ 
with a sufficiently small deviation 
with probability almost $1$.  
Let us consider the case for 
the canonical density operator
$\rho^{\mathrm{can}}(\beta,V,N):=\exp(-\beta H(V,N))/Z(\beta,V,N)$ where
$H(V,N)$ is the Hamiltonian and \\
$Z(\beta,V,N) := \mathrm{tr}[\exp(-\beta H(V,N))]$ is the partition function.
It can be shown that
$(\Delta \la\psi|\psi\ra)^2=Z(2\beta,V,N)/(Z(\beta,V,N))^2=
\exp(-2\beta(F(2\beta,V,N)-F(\beta,V,N)))$, where
$F(\beta,V,N)=-\beta^{-1}\ln Z(\beta,V,N)$ is the Helmholtz free energy. 
For thermodynamically sound systems, we have $F(\beta,V,N)=O(V)$
for $V \to \infty$, $\pd F(\beta,V,N)/\pd \beta$ $=\beta^{-2} S(\beta,V,N)$,
and the entropy $S(\beta,V,N)$ can be chosen to be positive 
for $0\le \beta <\infty$.  Therefore, we have
$(\Delta \la\psi|\psi\ra)^2\sim \exp[-f (\beta) V]$ with a
function $f(\beta)$ of $\beta $ satisfying $f(\beta)\ge 0$,
which implies that $\la \psi|\psi \ra=1$ is satisfied with
probability $1$ in the thermodynamic limit $V \to \infty$ for 
$0\le \beta <\infty$.  

When $\rho$ can be written in the form
$\rho=|J|^{-1}\sum_{j\in J}|j\ra\la j|$ where $J$ is a set with
the size $|J|<\infty$ and $\la j|j'\ra=\delta_{jj'}$, there is
another choice of PDF on the Hilbert space.  The PDF 
is given by the uniform distribution on the surface of hypersphere
$\la \psi|\psi\ra=1$ in the subspace $\mathcal{H}_J$ which is
spanned by $|j\ra (j\in J)$.  When all the thermodynamic variables
in $X$ are extensive variables such as the internal energy $U$, 
the microcanonical density operator 
$\rho^{\mathrm{mc}}(X)$
may be written in the above form with 
$\mathcal{H}_J$ being the subspace spanned by the energy
eigenvectors with the eigenvalues lying in $((1-\delta) U, U]$
where $0<\delta \ll 1$ and thermodynamic variables in $X$
other than $U$ (such as volume $V$ or particle number $N$) 
are fixed.  We denote the PDF by $P_{\mathrm{sp}}^{(X)}(\psi)$ where
the subscript `sp' stands for hyper`sp'here.  This type of
PDF is often used in the context of typicality
(e.g., Refs.~\cite{Sugita2006,Sugita2007,PopescuShortWinter2006}).  

In the present study, we consider the case such that
inverse temperature $\beta$ rather than internal energy $U$ is
used to specify the thermodynamic state and the corresponding
density operator is the canonical density operator, for which
the model PDF that is uniformly distributed on a hypersphere can
not be applied straightforwardly (but, see 
Refs.~\cite{SugiuraShimizu2012,SugiuraShimizu2013} for some attempts
to define an ensemble of pure states corresponding to canonical ensembles. )
Therefore, we use a normal distribution model PDF in the present study.  
Although it has the disadvantage that each pure state is not normalized in
the strict sense, it has the advantage that some properties of normal
distribution enable us to push forward the computation in 
Sec.~\ref{sec:boltzmannentropy}.  

\section{Quantum field systems}
\label{sec:quantumfield}
Let the spatial domain of the system be a $d$-dimensional cube 
with sides of length $L$ applied with periodic boundary conditions.  
The volume of domain is $V=L^d$.
Let the spatial coordinate be discretized with the unit length
$\Delta x$ in each direction and let $\mathcal K$ be 
the set of correspondingly discretized wavevectors
$\bm k=(n_1,\ldots,n_d) \Delta k$ where
$\Delta k=2\pi/L$ and
$n_j=-L/2\Delta x,-L/2\Delta x+1,\ldots,L/2\Delta x-1\ (j=1,\ldots,d)$.
The number of elements of $\mathcal K$, which is same as
the number of lattice points in the spatial domain,
is $V (\Delta x)^{-d}$.  
In the following, the thermodynamic limit 
will be taken by $V \to \infty$ with fixed 
$\Delta x$, or equivalently, $\Delta k \to 0$ with fixed 
$k_{\max}:=\sqrt{d}\pi/\Delta x$.  

We consider a bosonic or fermionic 
field in this domain.   
Annihilation and creation operators, $a_{\bm k}$ and $a^\dagger_{\bm k}$ 
respectively, associated with the wavevector $\bm k$ satisfy 
the commutation relation 
$[a_{\bm k},a^\dagger_{\bm k'}]=\delta_{\bm k,\bm k'}$ for bosonic fields, 
and the anticommutation relation 
$\{a_{\bm k},a^\dagger_{\bm k'}\}=\delta_{\bm k,\bm k'}$ for fermionic fields, 
where $\delta_{\bm k,\bm k'}$ is the Kronecker delta in the vector space, 
i.e., $\delta_{\bm k,\bm k'}=1$ for $\bm k=\bm k'$ and 
$\delta_{\bm k,\bm k'}=0$ otherwise.  
Let $\{n\}$ be a list of numbers $n_{\bm k} (\bm k \in \mathcal K)$ 
and $|\{n\}\ra$ be the Fock state,
\begin{equation}
|\{n\}\ra:={\prod_{\bm k}}'\frac{(a_{\bm k}^\dagger)^{n_{\bm k}}}{\sqrt{n_{\bm k}!}}|0\ra,
\label{eq:|n>}
\end{equation}
where $|0\ra$ is the vacuum specified by 
$a_{\bm k}|0\ra=0 (\bm k\in \mathcal K)$, 
$0 \le n_{\bm k} <\infty$ for bosonic fields and 
$n_{\bm k}=0,1$ for fermionic fields, and 
$\prod_{\bm k}'$ denotes an ordered multiplication according 
to an arbitrary rule.  Note that the rule is required to 
eliminate the ambiguity of the sign for the case of fermionic fields.  
The Fock states $|\{n\}\ra$ with $0 \le \sum_{\bm k}n_{\bm k} <\infty$ form 
an orthonormal basis of the Hilbert space $\mathcal H$ of the field 
system.  

We assume that the Hamiltonian $H$ of the field is 
invariant under the global phase translation
$a_{\bm k}\to a_{\bm k} \ee^{\ri \theta}$ 
for arbitrary $\theta \in \mathbb{R}$ and 
the spatial translation 
$a_{\bm k}\to a_{\bm k} \ee^{-\ri\bm k\cdot \delta \bm x}$ 
for arbitrary $\delta \bm x \in \mathbb{R}^{d}$. 
These imply that the Hamiltonian $H$ commutes with 
the particle number $\hat N:=\sum_{\bm k}a_{\bm k}^\dagger a_{\bm k}$ 
and the momentum $\bm p:=\sum_{\bm k}\bm k a_{\bm k}^\dagger a_{\bm k}$.  
The explicit form of the Hamiltonian up to fourth order 
in the field operators is given by, 
\begin{align}
  H&=\sum_{\bm k}\omega_{\bm k}a_{\bm k}^\dagger a_{\bm k}+
\frac{1}{2}
  \sum_{\bm k\bm q\bm r\bm s}g_{\bm k\bm q\bm r \bm s}\delta_{\bm k+\bm q,\bm r+\bm s}
  a_{\bm k}^\dagger a_{\bm q}^\dagger a_{\bm r} a_{\bm s}, 
\end{align}
where 
$\omega_{\bm k} \ge 0$ and $g_{\bm k\bm q\bm r\bm s}\in \mathbb{C}$.
In general, $H$ may contain higher-order terms in $a_{\bm k}$ and 
$a_{\bm k}^\dagger$.  
We assume that $\omega_{\bm k}, g_{\bm k\bm q\bm r\bm s}$ and
higher-order coefficients are prescribed in the domain of
continuous wavevector space and that they are continuous functions 
of the wavevectors.  
Let $|j\ra\ (j=0,1,2,\cdots)$ be simultaneous eigenstates of
the particle numbers $\hat N$, momentum $\bm p$, 
and the Hamiltonian $H$ with the eigenvalues $N_j$, $\bm p_j$ and $E_j$, 
respectively.  When some of the energy eigenstates are degenerate, i.e.,
$N_j=N_{j'}$, $\bm p_j=\bm p_{j'}$ and $E_j=E_{j'}$ for $j\ne j'$, 
we choose $|j\ra$ and $|j'\ra$ to be orthogonal 
so that all $|j\ra$'s form an orthonormal basis of the Hilbert space 
$\mathcal{H}$ of the field system.
The Fock states $|\{n\}\ra$ may be used as an orthonormal basis $|j\ra$ 
for free field systems, i.e.,
$H=\sum_{\bm k}\omega_{\bm k}a_{\bm k}^\dagger a_{\bm k}$, 
but the orthonormal basis $|j\ra$ does not coincide with
the Fock states $|\{n\}\ra$ for general (self-)interacting field systems.  

Hereafter, we consider the thermodynamic state $(\beta,V,N)$. 
Let $\mathcal{H}_N$ be the subspace of $\mathcal{H}$ spanned by
the eigenstates of $\hat N$ with the eigenvalue $N$.
We restrict the Hilbert space to $\mathcal{H}_N$ for 
the thermodynamic equilibrium state $(\beta,V,N)$ and 
the thermodynamic limit $V\to \infty$ will be
taken with fixed $N/V$.  

\section{Boltzmann entropy for quantum field systems}
\label{sec:boltzmannentropy}
Let the unitary operator $U_\Theta$ 
parametrized by $\Theta=(\theta_0,\theta_1,\theta_2,\ldots)\ 
(0\le \theta_j<2\pi )$ be given by 
\begin{equation}
U_\Theta |j\ra= \sum_j\ee^{\ri \theta_j}|j\ra. 
\label{eq:UTheta}
\end{equation}
%
Let $U_{\bm\kappa}$ be the unitary operator which shifts the field
in the wavevector space by a wavevector $\bm\kappa\in \mathcal K'$,
i.e., 
\begin{equation}
U_{\bm\kappa}a_{\bm k}U_{\bm\kappa}^\dagger=a_{\bm k+\bm \kappa}, 
\label{eq:Ukappa}
\end{equation}
where $\mathcal K'$ is a set of wavevectors $\bm\kappa$ satisfying 
$\bm\kappa=(n_1,\cdots,n_d)\Delta k$ with
$n_j=-\lfloor\zeta L/2\Delta x\rfloor,$
$-\lfloor\zeta L/2\Delta x\rfloor+1,\cdots,\lfloor\zeta L/2\Delta x\rfloor$
and $\zeta$ is a positive small parameter. 
Hereafter, we identify the wavevector 
$\bm k+ (2\pi/\Delta x) \bm \ell (\bm \ell \in\mathbb{Z}^d)$ with 
$\bm k$ so that all wavevectors under consideration belong to
$\mathcal K$.  Note that $U_{\bm\kappa}^\dagger=U_{-\bm\kappa}$.  
Provided that the vacuum $|0\ra$ is
invariant under the operation of $U_{\bm\kappa}$ for $\kappa\in\mathcal{K'}$,
the operator $U_{\bm\kappa}$ transfers a Fock state $|\{n\}\ra$ 
to another Fock state
$U_{\bm\kappa}|\{n\}\ra={\prod_{\bm k}}'
  (a_{\bm k+\bm\kappa}^\dagger)^{n_{\bm k}}/\sqrt{n_{\bm k}!}|0\ra$. 
In the thermodynamic limit, we put 
$\zeta=\zeta_0 (L/\Delta x)^{-\alpha}=\zeta_0 V^{-\alpha/d}(\Delta x)^\alpha$ with 
$0 < \alpha < 1$ and a positive small constant $\zeta_0$. 
Note that 
$\kappa_{\max}:=\sqrt{d}\lfloor \zeta L/2\Delta x\rfloor \Delta k$ scales 
as $\kappa_{\max}=O(V^{-\alpha/d})$ and 
the size of the set $\mathcal K'$ scales as 
$|\mathcal K'|\propto V^{1-\alpha}$. 

Let the 
operator $f(\psi)$ associated with 
a state $|\psi\ra$ be given by 
\begin{align}
f(\psi)&=\left(\prod_j\int_0^{2\pi}\frac{\rd\theta_j}{2\pi} \right)
\frac{1}{|\mathcal K'|}\sum_{\bm \kappa} U_\Theta U_{\bm \kappa} 
|\psi\ra\la\psi| U_{\bm \kappa}^\dagger U_\Theta^\dagger
\nonumber\\
&=\frac{1}{|\mathcal K'|}\sum_j\sum_{\bm \kappa}
\mathsf{P}_j U_{\bm \kappa} |\psi\ra\la\psi| U_{\bm \kappa}^\dagger \mathsf{P}_j, 
\label{eq:f_psi_quantum}
\end{align}
where $\mathsf{P}_j:=|j\ra\la j|$ is the projection operator and 
we used Eq.(\ref{eq:UAUdagger}) in the second equality.  
We propose to define the Boltzmann entropy $S(\psi)$ 
for quantum field systems as 
Eq.(\ref{eq:entropy_quantum}) with Eq.(\ref{eq:f_psi_quantum}).  
We will see that, under an assumption specified later, 
the proposed Boltzmann entropy $S(\psi)$ 
satisfies the properties (S1)--(S3) with respect to 
the ensemble of pure states given by the PDF 
$P_{\mathrm N}^{(\beta,V,N)}(\psi)$, that is 
the normal distribution model for 
the canonical density operator $\rho^{\textrm{can}}(\beta,V,N)$ 
associated with the thermodynamic state $(\beta,V,N)$.  
In the rest of this section, $(\beta,V,N)$ 
is denoted by $\gamma$ for convenience.

From (\ref{eq:f_psi_quantum}), we have 
\begin{align}
  \la j|f(\psi)|j'\ra &=f_j(\psi) \delta_{jj'},
\label{eq:rho_psi_delta}
\\
  f_j(\psi) &=
  \frac{1}{|\mathcal{K}'|}\sum_{\bm\kappa\in\mathcal{K}'}
  |\la j|U_{\bm \kappa}|\psi\ra|^2. 
\label{eq:rho_psi_j}
\end{align}
Let 
$H_{\bm\kappa}:=U_{\bm\kappa}H U_{\bm\kappa}^\dagger$ and
$\delta_{\bm\kappa}H:= H_{\bm\kappa}-H$.  The general expression
of $H_{\bm\kappa}$ up to the fourth order in the field operators
is given by 
\begin{align}
  H_{\bm\kappa}&=\sum_{\bm k}\omega_{\bm k-\bm\kappa}a_{\bm k}^\dagger a_{\bm k}
+\frac{1}{2}\sum_{\bm k\bm p\bm q\bm r}
g_{\bm k-\bm\kappa,\bm p-\bm\kappa,\bm q-\bm\kappa,\bm r-\bm\kappa}\delta_{\bm k+\bm p,\bm q+\bm r}
a_{\bm k}^\dagger a_{\bm p}^\dagger a_{\bm q} a_{\bm r}. 
\end{align}
Let $\rho^{\textrm{can}}_{\bm \kappa}(\gamma):=
U_{\bm\kappa}\rho^{\textrm{can}}(\gamma)U^\dagger_{\bm\kappa}=
\exp[-\beta(H+\delta_{\bm\kappa}H)]/Z(\gamma)$ and 
define an operator $A_{\bm\kappa}(\gamma)$ by 
\begin{equation}
\rho^{\textrm{can}}_{\bm\kappa}(\gamma)
=\rho^{\textrm{can}}(\gamma)(I + \lambda A_{\bm\kappa}(\gamma)),
\label{eq:Ak}
\end{equation}
where $I$ is the identity operator and $\lambda=1$ is 
a bookkeeping parameter. Since $\omega_{\bm k}$, 
$g_{\bm k\bm q\bm r\bm s}$ and the higher-order coefficients 
in $H$ are continuous in the wavevectors  $\bm k$, 
we have $A_{\bm\kappa}(\gamma)=O(\kappa)$ for 
$\kappa\to 0$ where $\kappa:=|\bm\kappa|$.  

Since we use the normal distribution model $P_{\mathrm{N}}^{(\gamma)}(\psi)$,
the coefficients $\psi_j=\la j|\psi\ra$ obey the multivariate
complex normal distribution with the means
$\overline{\psi_j}^{(\gamma)}=0$ and
the covariances
$\overline{\psi_j\psi_{j'}^*}^{(\gamma)}=\rho_j^{\mathrm{can}}(\gamma)\delta_{jj'}$ where 
$\rho_j^{\textrm{can}}(\gamma):=\la j|\rho^{\textrm{can}}(\gamma)|j\ra
=\exp[-\beta E_j]/Z(\gamma)$.  
From Eqs. (\ref{eq:rho_psi_j}) and (\ref{eq:Ak}), we have 
\begin{equation}
  \overline{f_j(\psi)}^{(\gamma)}=\rho^{\textrm{can}}_j(\gamma)
    \left(1+\frac{\lambda}{|\mathcal{K}'|}\sum_{{\bm\kappa}\in\mathcal{K}'}
\la j|A_{\bm\kappa}(\gamma)|j\ra\right).
\label{eq:rho_psi_bar}
\end{equation}
Let $\Delta f_j(\psi):=f_j(\psi)-\overline{f_j(\psi)}^{(\gamma)}$ 
be the fluctuation and
$C_{j_1j_2\cdots j_m}(\gamma):=$ \\
$\overline{\Delta f_{j_1}(\psi) \Delta f _{j_2}(\psi) \cdots \Delta f_{j_m}(\psi)}^{(\gamma)}$ be the associated $m$-th order moments.
In virtue of the multivariate complex normal distribution of $\psi_j$, 
$C_{j_1j_2\cdots j_m}(\gamma)$ can be expressed as 
\begin{align}
C_{j_1 j_2\cdots j_m}(\gamma)
&
=\frac{1}{|\mathcal{K}'|^m}
\sum_{\sigma\in\mathcal{P}}
\left(\prod_{m'=1}^m\sum_{\kappa_{m'}\in\mathcal{K}'}\right)
\left(
\prod_{m'=1}^m\la j_{m'}|
\rho^{\textrm{can}}_{\bm\kappa_{m'}}(\gamma) U_{\bm\kappa_{m'}-\bm\kappa_{\sigma(m')}}
|j_{\sigma(m')}\ra
\right), 
\label{eq:correlation}
\end{align}
where $\mathcal{P}$ is the set of permutations of $\{1,2,\ldots,m\}$ 
without fixed points.  
Note that $\la j|\rho^{\mathrm{can}}_{\bm\kappa}(\gamma)U_{\bm\kappa-\bm\kappa'}|j'\ra \ne 0$ only if $\bm p_j-N \bm\kappa =\bm p_{j'}-N \bm\kappa'$.  
For fixed $j$, $j'$ and $\bm\kappa$ there is at most one $\bm\kappa'$ 
that satisfies the above condition.  
There are $|\mathcal{K}'|^m=O(V^{m(1-\alpha)})$ terms in the summation 
$\prod_{m'=1}^m\sum_{\kappa_{m'}\in\mathcal{K}'}$, but at most 
$O(|\mathcal{K}'|^{m/2})=O(V^{m(1-\alpha)/2})$ terms are not $0$.  
The fact is essential in the following estimates.  

The Boltzmann entropy $S(\psi)$ depends on $\psi$ 
only through $f(\psi)$ as 
$S(\psi)=\sum_j S_1(f_j(\psi))$ where 
$S_1(x):=-x \ln x$.
Let us introduce bounding functions 
$S_1^{(\mathrm{u})}(x,y):=-y\ln y - (1+\ln y)(x-y)$ 
and $S_1^{(\mathrm{\ell})}(x,y):=S_1^{(\mathrm{u})}(x,y)
-y^{-1} (x-y)^2$, which are polynomial in $x$ and satisfy 
$S_1^{(\mathrm{\ell})}(x,y)\le S_1(x)\le S_1^{(\mathrm{u})}(x,y)$
for $x,y > 0$.
By substituting $x=f_j(\psi)$ and 
$y=\overline{f_j(\psi)}^{(\gamma)}$ into the inequalities, 
taking the average with respect to the thermodynamic state 
$\gamma$ and summing over $j$ yields 
\begin{align}
&\sum_j \left(S_1\left(\overline{f_j(\psi)}^{(\gamma)}\right) 
- \left(\overline{f_j(\psi)}^{(\gamma)}\right)^{-1} C_{jj}(\gamma)\right)
\le \overline{S(\psi)}^{(\gamma)}
\le \sum_j S_1\left(\overline{f_j(\psi)}^{(\gamma)}\right). 
\label{eq:S_psi_bounds}
\end{align}
The upper and lower bounds in the inequalities (\ref{eq:S_psi_bounds}) 
can be expanded in power series of $\lambda$ by using 
Eqs. (\ref{eq:Ak})--(\ref{eq:correlation}).  
Each $O(\lambda)$ term contains $A_{\bm\kappa}(\gamma)$ whose components 
$\la j|A_{\bm\kappa}(\gamma)|j'\ra$ are $O(\kappa)$ for fixed $V$.  
Here we employ the following assumption.
\begin{description}
\item{(A1)}
$O(\lambda)$ terms appearing in the estimate of 
$\overline{S(\psi)}^{(\gamma)}$ and $\overline{(\Delta S(\psi))^2}^{(\gamma)}$ 
can be neglected in the thermodynamic limit $V\to\infty$ with 
$\kappa_{\max}=O(V^{-\alpha/d})$.
\end{description}
Under assumption (A1), 
$\overline{f_j(\psi)}^{(\gamma)}$ and $C_{jj}(\gamma)$ in 
Eq. (\ref{eq:S_psi_bounds}) can be replaced by 
$\rho^{\textrm{can}}_j(\gamma)$ and 
$|\mathcal{K}'|^{-1}(\rho^{\textrm{can}}_j(\gamma))^2$ respectively.  
Thus, we arrive at 
\begin{equation}
\overline{S(\psi)}^{(\gamma)}=S(\gamma)+O(V^{-(1-\alpha)}), 
\label{eq:Spsi=Sgamma}
\end{equation}
which implies that (S1) is satisfied. 

For the estimate of $\overline{(\Delta S(\psi))^2}^{(\gamma)}$,
we introduce a function $\Delta S_1(x,y):=S_1(x)-S_1(y)$ and
a bounding function 
$\Delta S_1^{2(\mathrm{u})}(x,y):=((1-2\ln y)/2 y)^2 (x^2-y^2)^2$ 
such that 
$(\Delta S_1(x,y))^2 \le \Delta S_1^{2(\mathrm{u})}(x,y)$ for $x,y >0$. 
We have, 
\begin{align}
  \overline{(\Delta S(\psi))^2}^{(\gamma)}
  &=
  \sum_{jj'}\overline{
    \Delta S_1(f_j(\psi),\overline{f_j(\psi)}^{(\gamma)})
    \Delta S_1(f_{j'}(\psi),\overline{f_{j'}(\psi)}^{(\gamma)})
  }^{(\gamma)}
  \nonumber\\
  &\quad -\left(
  \sum_j \Delta S_1(f_j(\psi),\overline{f_j(\psi)}^{(\gamma)})
  \right)^2,
  \nonumber\\
  &\le \left(
  \sum_j\left(\overline{
    \Delta S_1^{2(\mathrm{u})}(f_j(\psi),\overline{f_j(\psi)}^{(\gamma)})
  }^{(\gamma)}\right)^{1/2}
  \right)^2,
  \label{eq:boundDSpsi2}
\end{align}
where we used the Cauchy-Schwarz inequality and the bounding inequality
regarding $\Delta S_1^{2(\mathrm{u})}(x,y)$ in the last inequality.
The right-hand side of inequality (\ref{eq:boundDSpsi2}) 
can be written in terms of the moments
$C_{j_1j_2\cdots j_m}(\gamma)$ in Eq.(\ref{eq:correlation}).  
By applying assumption (A1) to inequality (\ref{eq:boundDSpsi2}), 
we can show that 
\begin{equation}
\frac{\overline{(\Delta S(\psi))^2}^{(\gamma)}}{S(\gamma)^2}=O(V^{-(1-\alpha)}),
\end{equation}
which implies that (S2) is satisfied. 

Regarding the dynamics, we have
\begin{equation}
  f_j(\psi(t))=\frac{1}{|\mathcal{K}'|}
  \sum_{\bm\kappa\in\mathcal{K}'}
  \left|
  \sum_{j'} \la j|U_{\bm\kappa}|j' \ra \psi_{j'}(0) \ee^{-\ri E_{j'} t}
  \right|^2.
\label{eq:f_j_psi_t}
\end{equation}
In the case of free field systems, $|j\ra$ and $|j'\ra$ are
Fock states and $U_{\bm\kappa}^\dagger|j\ra=U_{-\bm\kappa}|j\ra$
is also a Fock state.
Since the Fock states form an orthonormal basis, 
$\la j|U_{\bm\kappa}|j'\ra \ne 0$ is satisfied for at most one 
$j'$ for arbitrary fixed $j$ and $\bm\kappa$.
This implies that $f_j(\psi(t))$ is independent of $t$
for free field systems.  For general (self-)interacting field systems,
we have $[H,H_{-\bm \kappa}]\ne 0$ for $\bm \kappa\ne\bm 0$.  
Note that $U_{-\bm\kappa}|j\ra (j=1,2,\cdots)$ are eigenstates 
of $H_{-\bm\kappa}$, and that there are some $j$ such that 
$\la j|U_{\bm\kappa}|j'\ra\ne 0$ for two or more $j'$'s.  
The element $f_j(\psi)$ depends on $t$ when $\psi_{j'}(0)\ne 0$ 
for those $j'$'s.  
Furthermore, the $t$-dependence of $f_j(\psi(t))$ leads to 
the $t$-dependence of $S(\psi(t))$.  Thus, (S3) is satisfied
for (self-)interacting field systems.  

\section{Discussions}
\label{sec:discussions}
A different and simpler definition of the Boltzmann entropy 
can be given by substituting 
\begin{align}
f_{\mathrm{d}}(\psi)
&=\left(\prod_{j=1}^{\mathcal D}\int_0^{2\pi}\frac{\rd\theta_j}{2\pi} \right)
U_\Theta |\psi\ra\la\psi| U_\Theta^\dagger, 
\nonumber\\
&=\sum_j
\mathsf{P}_j |\psi\ra\la\psi| \mathsf{P}_j, 
\label{eq:rho_d}
\end{align}
into $f(\psi)$ of Eq. (\ref{eq:entropy_quantum}).  
This type of entropy, denoted $S_{\mathrm{d}}(\psi)$, is often called
the diagonal entropy (see, e.g., Ref. \cite{Polkovnikov2011}). 
It can be shown 
that $S_{\mathrm{d}}(\psi)$ satisfies (S1) and (S2) but fails to 
satisfy (S3), i.e., $S_{\mathrm{d}}(\psi(t))$ do not depend on $t$ 
under the Hamiltonian dynamics.
In order that $f(\psi(t))$ and $S(\psi(t))$ to depend on time, 
a parametrized set of unitary operators which do not commute 
with the Hamiltonian $H$ is necessary.
The situation is conceptually akin to selecting a decomposition of
the Hilbert space $\mathcal{H}$ whose projection operators $\mathsf{P}_\nu$
do not commute with $H$ in VN29. 
Furthermore, the introduction of the set of unitary operators
should not violate (S1) and (S2).
Following the preceding study on the Boltzmann entropy for 
classical field systems \cite{Yoshida2015}, we introduced
the set of operators $U_{\bm\kappa} (\bm\kappa \in \mathcal{K}')$
that shifts the field in the wavevector space to fulfill
the above requirements.  The underlying idea of the operator
$U_{\bm\kappa}$ may be given as follows.  
When the Hamiltonian $H$ is invariant under the spatial translation
and global phase translation, 
the correlation in the wavevector space takes the form 
$\mathrm{tr}[a_{\bm k}^\dagger a_{\bm k'} \rho^{\mathrm{can}}(\beta,V,N)]=
G_{\bm k}\delta_{\bm k,\bm k'},
\mathrm{tr}[a_{\bm k} a_{\bm k'} \rho^{\mathrm{can}}(\beta,V,N)]=
\mathrm{tr}[a_{\bm k}^\dagger a_{\bm k'}^\dagger \rho^{\mathrm{can}}(\beta,V,N)]=0$.  
This implies that the different 
wavevector modes are uncorrelated at the level of 
the second-order moments.  When $G_{\bm k}$ is a smooth function 
of $\bm k$, which may be expected when $\omega_{\bm k}$ 
and higher-order coefficients in the Hamiltonian $H$ are smooth, 
the neighboring wavevector modes $\bm k$ and 
$\bm k+\bm \kappa (\kappa\in\mathcal{K}')$ may be interpreted 
as statistically quasi-independent replicas of each other. 
The interacting wavevector modes
$a_{\bm k_1},\ldots,a_{\bm k_m},a_{\bm k'_1}^\dagger, \ldots, a_{\bm k'_m}^\dagger$
satisfy the momentum preserving condition
$\sum_{m'=1}^m\bm k_{m'}-\sum_{m'=1}^m\bm k'_{m'}=\bm 0$ and
the condition is not violated by a shift in the wavevector space,  
$\bm k \to \bm k + \bm \kappa$.  
Based on these considerations, it may be appropriate to regard
that the states $|\psi\ra$ and $U_{\bm \kappa}|\psi\ra$
($\kappa \in\mathcal{K}'$) ``resemble'' each other.  

In the present study, we chose the set of thermodynamic 
variables $(\beta,V,N)$ to specify a thermodynamic state 
and the corresponding canonical density operator 
$\rho^{\mathrm{can}}(\beta,V,N)$ is interpreted as an ensemble 
of unnormalized pure states $P_{\mathrm{N}}^{(\beta,V,N)}(\psi)$ 
using the normal distribution model.  
Exploiting this setting of the ensemble of pure states, 
we showed, with an additional assumption (A1), that 
(S1)--(S3) hold.  
When the set of thermodynamic variables $(U,V,N)$, 
where $U$ is the internal energy, is chosen, 
the density operator may be given by
that of the microcanonical ensemble  
$\rho^{\mathrm{mc}}(U,V,N)=|J_U|^{-1}\sum_{j\in J_U}|j\ra\la j|$
where $J_U$ is the set of indices $j$ satisfying 
$(1-\delta) U < E_j \le U$ with $H|j\ra=E_j|j\ra$ and 
$0<\delta\ll 1$, and $|J_U|$ is the size of the set $J_U$.  
As mentioned in Sec.~\ref{sec:normaldistributionmodel}, 
the model PDF $P_{\mathrm{sp}}^{(U,V,N)}$ which is distributed uniformly
on the surface of a hypersphere may be used in this case. 
%
%
The PDF $P_{\mathrm{N}}^{(\beta,V,N)}(\psi)$ approximates 
$P_{\mathrm{sp}}^{(U(\beta,V,N),V,N)}(\psi)$ in the sense that 
$(\Delta\la \psi|\psi \ra) \to 0$ for $V \to \infty$ and 
that 
$\Delta U(\beta,V,N)/U(\beta,V,N)=O(V^{-1/2})$,
where $\Delta U(\beta,V,N)):=(\mathrm{tr}[\rho^{\mathrm{can}}(\beta,V,N)
  (H-U(\beta,V,N))^2])^{1/2}$, 
holds for thermodynamically sound systems.  
Taking into account the resemblance of the two PDFs 
$P_{\mathrm{N}}^{(\beta,V,N)}(\psi)$ and 
$P_{\mathrm{sp}}^{(U(\beta,V,N),V,N)}(\psi)$, 
it is probable that the properties (S1)--(S3) hold not only
for $P_{\mathrm{N}}^{(\beta,V,N)}(\psi)$ 
but also for $P_{\mathrm{sp}}^{(U,V,N)}(\psi)$. 
If (S1)--(S3) hold for $P_{\mathrm{sp}}^{(U,V,N)}(\psi)$,
$\Delta S(\psi):=S(\psi)-S(U=\la\psi|H|\psi\ra,V,N)$ would
be almost $0$ for almost all states $|\psi\ra$ with respect to
the PDF $P_{\mathrm{sp}}^{(U,V,N)}(\psi)$.  We interpret 
these states as thermodynamic equilibrium states.  Then, 
$\Delta S(\psi)$ gives a measure of departure from
the thermodynamic equilibrium.  States with large
$|\Delta S(\psi)|$ correspond to nonequilibrium states.
We hope that $\Delta S(\psi) < 0$ for most nonequilibrium states and
that the thermalization process corresponds to
$\lim_{t\to \infty} \Delta S(\psi(t)) = 0$ with
$\rd S(\psi(t))/\rd t >0$.  

The eigenstate thermalization hypothesis (ETH) 
\cite{Deutsch1991,Srednicki1994} states that a single energy 
eigenstate can behave as a thermal state.  In the present context, 
the ETH is given as $\Delta S(\psi)\approx 0 $ for almost all
energy eigenstates $|\psi\ra =|j\ra $.  One can see from 
Eqs. (\ref{eq:entropy_quantum}), 
(\ref{eq:f_psi_quantum})--(\ref{eq:rho_psi_j}) 
that the expression of the operators 
$U_{\bm \kappa} (\kappa\in \mathcal{K}')$ in the energy eigenstate basis 
$|j\ra$, that is $\la j'|U_{\bm k}|j \ra$, is required to compute 
$S(j):=\left. S(\psi)\right|_{|\psi\ra=|j\ra}$.  
In the case of a free field system, there is only one $j'$ such that 
$\la j'|U_{\bm k}|j \ra \ne 0$ for every fixed $j$ and we have 
$S(j)=-\ln |\mathcal{K}'|\propto -(1-\alpha)\ln V = o(V)$ and 
$S(j)\ne S(U=E_j,V,N)=O(V)$, which implies that the ETH is not valid
for free fields.  
For general fields with interactions, there can be more than one
$j'$ such that $\la j'|U_{\bm k}|j \ra\ne 0$ for every fixed $j$.
If there are sufficiently many such $j'$s, then there is a possibility 
that the ETH is valid for some fields with interactions.  
When the ETH is valid, nonequilibrium states $|\psi\ra$, if they exist, 
are linear superpositions of energy eigenstates that are 
thermodynamic equilibrium states.  
Let us fix an initial state as
$|\psi(0)\ra=\sum_j\psi_{j}(0)|j\ra$.  
Only when $\psi_{j'}(0)$ for various $j'$ are in some special 
coherent relations between each other, 
$f_j(\psi(t))$ in Eq. (\ref{eq:f_j_psi_t}) with $t=0$ would take
an atypical value.  
And when $f_j(\psi(0))$ takes an atypical value for sufficiently many $j$'s,
$\Delta S(\psi(0))$ would deviate significantly from $0$, i.e., 
$|\psi(0)\ra$ is a nonequilibrium state.  Even when $|\psi(0)\ra$
is a nonequilibrium state, the factors $\ee^{-\ri E_j' t}$ in 
Eq. (\ref{eq:f_j_psi_t}) destroy the coherence for $t> \epsilon$
with some small $\epsilon (>0)$ and $\Delta S(\psi(t))\approx 0$
would be achieved, which implies the thermalization.  
The present dynamical description of thermalization is consistent
with that depicted in Figure 2 of Ref.\cite{RigolDunjkoOlshanii2008}.  

In \v{S}DA19, a coarse-graining in position space is introduced to 
construct a Boltzmann entropy for systems of many quantum particles.  
Let the spatial domain of the system $\mathcal{D}$ be divided into 
disjoint subdomains $\mathcal{D}_\ell(\ell=1,2,\cdots)$ corresponding 
to the coarse-graining.  
Let $\hat n_\ell$ be the number operator of the particles 
in the subdomain $\mathcal{D}_\ell$ and  $\mathcal{H}_{\nu}$ 
be the subspace of the Hilbert space $\mathcal{H}$ that is spanned 
by the simultaneous eigenstates of $\{\hat n_1,\hat n_2,\cdots\}$ with 
the eigenvalues $\nu=\{n_1,n_2,\cdots\}$.  
Let $\mathsf{Q}_\nu$ be
the projection operator on $\mathcal{H}_{\nu}$ and 
$\mathsf{P}_j=|j\ra\la j|$ where $|j\ra (j=0,1,2,\cdots)$ are
energy eigenstates.  
The observational entropy $S_{xE}(\psi)$ with 
coarse-graining in position space and fine-graining in energy  
in the terminology of \v{S}DA19 is formulate by 
\begin{equation}
  S_{xE}(\psi)=-\sum_{\nu,j}
  \mathrm{tr}[\mathsf{P}_j\mathsf{Q}_\nu|\psi\ra\la\psi|
    \mathsf{Q}_\nu\mathsf{P}_j]
\ln
\frac{\mathrm{tr}[\mathsf{P}_j\mathsf{Q}_\nu |\psi\ra\la\psi|
    \mathsf{Q}_\nu\mathsf{P}_j]}
     {\mathrm{tr}[\mathsf{P}_j\mathsf{Q}_\nu\mathsf{P}_j]}, 
\end{equation}
in analogy with Eq.(\ref{eq:S_decom}).
Note that $S_{xE}$ in \v{S}DA19 
is defined for the general density matrix $\rho$ but here we restrict 
it to be a function of a pure state $\rho=|\psi\ra\la \psi| $.  
Since $\mathsf{P}_j$ and
$\mathsf{Q}_\nu$ do not commute in general, the density operator 
$f_{xE}(\psi)$ associated with the pure state $\psi$ which satisfies 
$S_{xE}(\psi)=-\mathrm{tr}[f_{xE}(\psi)\ln f_{xE}(\psi)]$ 
can not be formulated in general.  Thus, 
$S_{xE}(\psi)$ does not fit into the formalism given in 
Sec.{\ref{sec:formalism}} in a strict sense.  
However, we may discuss conceptional similarity and difference
between the Boltzmann entropy $S(\psi)$, 
Eq.(\ref{eq:entropy_quantum}) with Eq.(\ref{eq:f_psi_quantum}), 
proposed in the present paper and $S_{xE}(\psi)$ in \v{S}DA19.
They are similar in the sense that they both use $U_{\Theta}$,
$\mathsf{P}_j$, or the fine-graining in energy in the terminology
of \v{S}DA19.  The difference between the two 
Boltzmann entropies is that the shift in wavevector space is used in 
$S(\psi)$ while the coarse-graining in position space is used in 
$S_{xE}(\psi)$.  The coarse-graining method is supposed to work well 
when the interaction between the particles is collisional or short ranged, 
so that $S_{xE}(\psi)$ would not be appropriate when the interaction is 
long-ranged beyond the coarse-graining scale.  The Boltzmann entropy 
$S(\psi)$ in the present study has no limitation in the interacting scale 
range.  On the other hand, the method of shift in the wavevector space 
requires the invariance of the Hamiltonian under the spatial translation, 
so that $S_{xE}(\psi)$ may be more suitable in the cases such that 
an external potential field depending on position is present.  
Many other definitions of Boltzmann entropy satisfying (S1)--(S3) 
would be possible.  Actually, the examples of Boltzmann entropy 
other than $S_{xE}(\psi)$ are given in \v{S}DA19. 
Among many definitions, an appropriate one should be 
used depending on the situation.  

The main purpose of this paper is to propose the Boltzmann entropy
$S(\psi)$ for quantum field systems given by 
Eq.(\ref{eq:entropy_quantum}) with Eq.(\ref{eq:f_psi_quantum}) and
to show its potential for characterizing nonequilibrium dynamical
processes including thermalization.  Validation of the assumption (A1)
and detailed analysis of $\la j'|U_{\bm k}|j\ra$ and ETH for some specific
field systems are beyond the scope of the present study.  They may be 
left for future studies.   

\section*{Acknowledgments}
The author thanks Yasuhiro Tokura for valuable discussions.  


\end{document}